\documentclass[fleqn,usenatbib]{mnras}

\usepackage{newtxtext,newtxmath}

\usepackage[T1]{fontenc}

\DeclareRobustCommand{\VAN}[3]{#2}
\let\VANthebibliography\thebibliography
\def\thebibliography{\DeclareRobustCommand{\VAN}[3]{##3}\VANthebibliography}

\usepackage{graphicx}	
\usepackage{amsmath}	
\usepackage{url}
\usepackage{hyperref}
\usepackage{lscape}
\usepackage{caption}

\title[Foot points of the Cool loops]{Spectroscopic Diagnostic of the Footpoints of the Cool loops}

\author[B. Suresh Babu et al.]{
B. Suresh Babu$^{1}$,
Pradeep Kayshap$^{1}$\thanks{E-mail: virat.com@gmail.com},
Sharad C. Tripathi$^{1}$,
P. Jel\'\i nek$^{2}$
and B.N. Dwivedi$^{3}$
\\
$^{1}$School of Advanced Sciences and Languages, VIT Bhopal University, Kothrikalan, Sehore-466114, Madhya Pradesh, India\\
$^{2}$University of South Bohemia, Faculty of Science, Department of Physics, Brani\v sovsk\'a 1760, CZ -- 370 05 \v{C}esk\'e Bud\v{e}jovice, Czech Republic\\
$^{3}$RGIPT, Jais Amethi-229304, India\\
}

\date{Accepted XXX. Received YYY; in original form ZZZ}

\pubyear{2023}

\begin{document}
\label{firstpage}
\pagerange{\pageref{firstpage}--\pageref{lastpage}}
\maketitle
\begin{abstract}
Statistically, the cool loop's footpoints are diagnosed using Si~{\sc iv} resonance lines observations provided by Interface Region Imaging Spectrograph (IRIS). The intensity and Full Width at Half Maximum (FWHM) of the loop's footpoints in $\beta${--}$\gamma$ active regions (ARs) are higher than the corresponding parameters of footpoints in $\beta$ ARs. However, the Doppler velocity of footpoints in both ARs are almost similar to each other. The intensities of footpoints from $\beta${--}$\gamma$ AR is found to be around 9 times that of $\beta$ AR when both ARs are observed nearly at the same time. The same intensity difference reduces nearly to half (4 times) when considering all ARs observed over 9 years. Hence, the instrument degradation affects comparative intensity analysis. We find that Doppler velocity and FWHM are well-correlated while peak intensity is neither correlated with Doppler velocity nor FWHM. The loop's footpoints in $\beta$-$\gamma$ ARs have around four times more complex Si~{\sc iv} spectral profiles than that of $\beta$ ARs. The intensity ratios (Si~{\sc iv} 1393.78~{\AA}/1402.77~{\AA}) of the significant locations of footpoints differ, marginally, (i.e., either less than 1.9 or greater than 2.10) from the theoretical ratio of 2, i.e., 52\% (55\%) locations in $\beta$ ($\beta${--}$\gamma$) ARs significantly deviate from 2. Hence, we say that more than half of the footpoint locations are either affected by the opacity or resonance scattering. 
We conclude that the nature and attributes of the footpoints of the cool loops in $\beta$-$\gamma$ ARs are significantly different from those in $\beta$ ARs.


\end{abstract}


\section{Introduction}
Active regions (ARs) are the areas with strong magnetic fields, and they encompass the curvilinear magnetic flux confining plasma (see \citealt[and references cited therein]{FR2010}) referred to as coronal loops (e.g., \citealt{DelZanna2003, FR2010}). Coronal loops include hot core loops (\citealt{DelZanna2008}), warm loops (\citealt{DelZanna2011}), and fan loops (\citealt{Winebarger2013, Young2012, Warren2009}). Here, it is worth mentioning that different types of coronal loops are well resolved and studied using observations taken by various instruments (e.g.,\citealt{Ugarte-Urra_2009, FR2010, DT2012, Girijesh2022}). In addition to the coronal loops, ARs do have another type of loop structure which are visible at low temperature (i.e., transition region (TR)). They are referred to as cool loops or TR loops (e.g., \citealt{Hansteen2014, Huang2015, Polito2016, Rao2019a, Rao2019b}).These cool loops are found to be well-structured and are dynamic structures of the TR. Hence, the diagnosis of the cool loops is extremely important to understand the TR and the link between the corona from the lower atmosphere in a better way. Unfortunately, cool loops are not examined in depth, so far, due to the lack of high-resolution observations. Due to the less availability of studies on cool loops our understanding about them is limited.\\
In several studies of cool loops, an off-limb loop was investigated using coronal diagnostic spectrometer observations (CDS; \citealt{Harrison1995}), and the Doppler velocity of around 50 km/s was found along the loop (\citealt{Brekke1997}). Similarly, \citealt{Chae2000} had investigated an off-limb cool loop using observations from Solar Ultraviolet Measurements of Emitted Radiation (SUMER) instrument (\citealt{Wilhelm1995}) onboard Solar and Heliospheric Observatory (SoHO; \citealt{Domingo1995}). Furthermore, using SUMER observations, \cite{Doyle2006} have reported net redshifts of $\approx$20 km/s in footpoint of cool loops using N~{\sc v} 1238.82~{\AA}.\\

The Interface Region Imaging Spectrograph (IRIS; \citealt{DePon2014}) provides an opportunity to diagnose the cool loops in great detail as this instrument observes TR with unprecedented spatial and spectral resolution. Some research on cool loops has already been conducted utilising IRIS observations. Mostly, it is reported that the plasma flows up along the loop from one footpoint (i.e., blueshifted footpoint) and falls down towards another footpoint of the loop (i.e., redshifted footpoint) {--} siphon flows (i.e., driven by the pressure difference of the plasma at the footpoints of the loops having different magnetic field strengths) in the AR loops (\citealt{Doyle2006, Bethge2012}). 

As a result, one footpoint of a cool loop can be red-shifted while the other is blue-shifted, or vice versa. Firstly, using the IRIS observations, \cite{Huang2015} showed the occurrence of siphon flow in cool loops, and they found significant variations in the blue-shifted (i.e., from -5 to -10 km/s)  and red-shifted footpoints (i.e., 12 to 20 km/s) of the cool loops in Si~{\sc iv} line. Similar to \cite{Huang2015}, using the same Si~{\sc iv}, \cite{Rao2019b} have also reported significant variations in the blueshifts (i.e., 5{--}10 km/s) as well as redshifts (i.e., 10{--} 15 km/s) in footpoints of the cool loops. Furthermore, these blue-shifted and red-shifted footpoints were investigated at different heights using different IRIS spectral lines (\citealt{Rao2019b}), and it was found that the blueshifts as well as redshifts at the footpoints of the loops increase from the photosphere to the transition region. Apart from the siphon flow of cool loops, the TR is dominated by redshifts, for instance, \cite{Brekke1997} have shown that TR of quiet-Sun (QS) has redshifts of 5 km/s. Similarly, some other works also report the dominance of redshifts in QS, coronal hole (CH), and TR of ARs (e.g., \citealt{PJ1999, Dadashi2011, Dadashi2012, Kayshap2015}).\\ 

The cool loops are the integral feature of the solar TR, and most of the spectral profiles forming within the TR are non-Gaussian in nature. Using Solar Ultraviolet Measurements of Emitted Radiation (SUMER) onboard Solar and Heliospheric Observatory (SoHO) observations, it has been found that shown that several TR lines (e.g., C~{\sc ii} 1335~{\AA}, C~{\sc iii} 977~{\AA}, Si~{\sc iv} 1403~{\AA}, and many more) are non-Gaussian, and the spectral lines are well characterized by double Gaussian for core and broad second components (e.g., \citealt{Peter1999, Peter2000, Peter2001, Peter2006}). Further, using Extreme-ultraviolet Imaging Spectrometer (EIS)/Hinode observations, \cite{Peter2010} has shown that not only the TR lines but also coronal spectral lines of the active region (AR) are non-Gaussian, i.e., the spectral lines can be well fitted with double Gaussian. Recently, with the help of the IRIS observations, the Si~{\sc iv} in the ARs shows excessive wing emission (i.e., non-Gaussian profile), and it can be well fitted with non-Maxwellian $\kappa$ distribution (\citealt{Dudik2017}). Apart from these various solar regions (e.g., network, inter-network, and ARs), the non-Gaussian profiles do exist in the network jets (\citealt{Kayshap2021}). Hence, in short, we can say that the nature of the spectral profiles in TR is important and may reveal important physical processes in the in-situ plasma.\\

There are different types of AR depending on the complexity, namely, unipolar, bipolar, and complex groups (for more details see \citealt{Hale1919}). The cool loops exist in all ARs irrespective of their type. However, the physical properties of cool loops may change depending on the complexity of ARs. Hence, the present work is an attempt to diagnose the footpoints of the cool loops in two different types of AR, i.e., $\beta$ and $\beta$-$\gamma$. We have found that the physical properties/behaviour of the footpoints of cool loops of $\beta$ type ARs are significantly different from that of $\beta$-$\gamma$ type ARs. The paper is organized as follows: Section~\ref{section:obs_detail} describes the observational details and data reductions. The results are described in Section~\ref{section:results} followed by a summary and conclusion in the last Section.  



\section{Observations and data analysis}
\label{section:obs_detail}
The Interface Region Imaging Spectrograph (IRIS) provides high-resolution spectroscopic observations of various regions of the solar atmosphere including AR. IRIS observes far-ultraviolet (FUV) and near-ultraviolet (NUV) parts of the solar spectrum (see \citealt{DePon2014}). The FUV, as well as NUV spectra have several emission and absorption lines, and these lines are important to diagnose the different regions of the solar atmosphere. Si~{\sc iv} spectral lines (i.e., Si~{\sc iv} 1393.755~{\AA} and Si~{\sc iv} 1402.77~{\AA}) form in TR, and the cool loops are the main features of TR. We have utilized both Si~{\sc iv} lines to diagnose the foot points of the cool loops. In this work, we have utilized 10 different AR observations to diagnose the foot points of the cool loops.\\
\begin{figure*}
	\includegraphics[trim=4.0cm 0.2cm 4.0cm 0.0cm, scale=1.0]{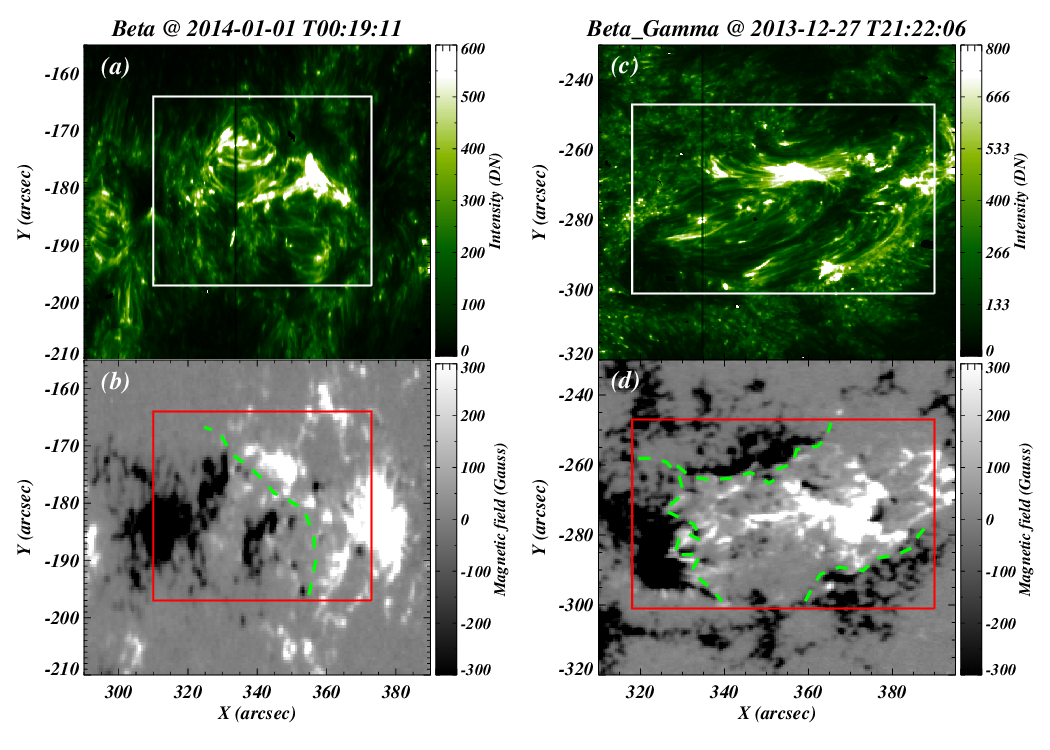}
    \caption{The panel (a) and panel (c) display IRIS/SJI 1400~{\AA} map of $\beta$ type AR11937 (i.e., 1$^{st}$ observation in table~\ref{tab:obs_details}) and first $\beta$-$\gamma$ type AR11934 (i.e., 6$^{th}$ observation in table~\ref{tab:obs_details}). The overplotted white rectangular boxes are showing the regions of interest (ROI). Similarly, the panels (b) and (d) display the LOS magnetogram of AR11937 and AR11934, respectively. Here, the same ROI is displayed by red rectangular boxes. The green dashed lines on the panels (b) and (d) displays the polarity inversion lines.} 
    \label{fig:fig1}
\end{figure*}
In Figure~\ref{fig:fig1}, we have displayed IRIS/SJI 1400~{\AA} intensity maps of AR11937 (first $\beta$ type; (panel a)) and AR11934 (first $\beta$-$\gamma$ type; panel c). The first $\beta$ type AR11937 (first observation of the table~\ref{tab:obs_details}) and the first $\beta$-$\gamma$ type AR 11934 (sixth observation of the table~\ref{tab:obs_details}) were observed on January 1$^{st}$, 2014 and December 27$^{th}$, 2013, respectively. The overplotted white boxes in panels (a) and (c) outline the region of interest (ROI) in both types of AR and ROI is filled with several cool loops. \\

Further, we have also displayed the Heliosesmic and Magnetic Imager (HMI) line-of-sight (LOS) magnetogram of AR11937 (panel b) and AR11934 (panel d). In both LOS magnetograms, the black region corresponds to negative polarity while the white region corresponds to positive polarity. Here, please note that the red rectangular boxes on these magnetograms display the same ROI as shown by white boxes in panels a \& c. In case of AR11937 (i.e., panel b), we have drawn the polarity inversion line (PIL) by green dashed line that clearly distinguish the positive and negative polarity, and one PIL is sufficient to separate positive and negative polarity. The existence of only one PIL justifies that AR11937 has a simple bipolar nature, and it is categorized as $\beta$ type AR. On the contrary, at least, we need three different PILs (see three green dashed lines within the red box in panel d) to draw the boundary between positive and negative polarity in the magnetogram of AR11934. Hence, unlike the AR11937, the presence of three different PILs justify that this AR11934 has a complex configuration of the magnetic field. Therefore, AR11934 is a $\beta$-$\gamma$ type AR.  Later, we checked the LOS magnetograms of all ARs, and we found that five out of 10 observations are $\beta$ type ARs while the rest five observations are of $\beta${--}$\gamma$ type AR (see table~\ref{tab:obs_details}). All the necessary details of all $\beta$ and $\beta${--}$\gamma$ type ARs are tabulated in the table\ref{tab:obs_details}. Further, please note that the information about the types of ARs are mentioned on the solar monitor website\footnote{\href{https://www.solarmonitor.org/index.php}{www.SolarMonitor.org}}, and our classification is consistent with the information provided on the solar monitor website.\\

\begin{table*}
	\centering
	\caption{This table describes all the necessary details of the IRIS active-region (AR) observations, which are utilized in the present work.}
	\label{tab:obs_details}
	\begin{tabular}{lcccccr} 
		\hline
    Sr. No. &  Type of AR & AR Number & Date $\&$ Time & Exposure/Cadence Time (s) & Central coordinates (x,y) & $\mu$ \\
		\hline
  
		1 & Beta & AR11937 & 2014-01-01 00:04:31-00:39:13 & 3.99/5.2 & 343", -237" & 0.900\\
  \hline
		2 & Beta & AR12641 & 2017-03-02 12:31:35-14:21:43 & 3.99/5.2 & -179", 373" & 0.902\\
  \hline
		3 & Beta & AR12832 & 2021-06-11 13:09:55-13:58:52 & 7.99/9.2 & 295", 235" & 0.919\\
  \hline
  	    4 & Beta & AR12458 & 2015-11-26 01:37:27-02:27:27 & 7.99/9.4 & -559", 138" & 0.800\\
  \hline
  	  5 & Beta & AR12692 & 2017-12-23 00:40:21-01:29:27 & 7.99/9.2 & -290", 343" & 0.883\\
  \hline
            6 & Beta-Gamma & AR11934 & 2013-12-27 21:02:38-21:36:29 & 3.99/5.1 & 324",-270" & 0.898\\
  \hline
    	  7 & Beta-Gamma & AR12712 & 2018-06-01 00:05:42-01:32:07 & 14.99/16.2 & 299",279" & 0.904\\
  \hline
    	  8 & Beta-Gamma & AR13226 & 2023-02-15 13:11:59-14:38:23 & 14.99/16.2 &  -234",281" & 0.924\\
  \hline
            9 & Beta-Gamma & AR10317 & 2022-05-24 05:36:48-06:25:45 & 7.99/9.2 & 521",235" & 0.803\\
  \hline
    	  10 & Beta-Gamma & AR13180 & 2023-01-04 08:41:24-09:30:30 & 7.99/9.2 & -223",380" & 0.888\\
  \hline
	\end{tabular}
\end{table*}
We have utilized uncalibrated (i.e., data numbers (DN/s)) as well as calibrated spectra (i.e., erg/cm$^{2}$/s/sr) in the present work. We have used iris$\_$getwindata.pro routine with calib keyword to get calibrated spectra. It is to be noted that we have used the latest version (version - 009) of iris calibration to get the calibrated spectra. The same routine is used with the norm keyword (and without calib keyword) to get uncalibrated spectra. In the next step, Si~{\sc iv} 1402.77~{\AA} line is fitted with the single Gaussian function to get the peak (or central) intensity, centroid, and Gaussian width (i.e., $\sigma$) at each location of ROI. It is also to be noted that Gaussian fit is applied on both uncalibrated and calibrated spectral profiles. Then, using peak intensity and Gaussian width, we have calculated the total intensity (i.e., the area under the curve) at each location of ROI. Again, it is to be noted that the total intensity is estimated in DN/s as well as erg/cm$^{2}$/s/sr. Further, the Gaussian width is converted into full width at half maximum (FWHM) using FWHM = 2.3548$\times$Gaussian width, and the centroid is converted into the Doppler velocity with the help of the rest wavelength of Si~{\sc iv} line. Here, it should be noted that estimating the rest wavelength is crucial as it can directly affect the Doppler velocity. There are three different methods to calculate the rest wavelength, namely, limb method, chromospheric method, and lamp method (\citealt{PJ1999}).\\

The present work utilizes the neutral/single ionized lines (i.e., cool lines) to estimate the rest wavelength of the Si~{\sc iv} line. Usually, the cool photospheric/chromospheric lines have low systematic Doppler shifts, and this low systematic Doppler shift is considered the least plasma flow that exists in the solar atmosphere. Therefore, cool photospheric/chromospheric lines are used in this method to calibrate the rest wavelength of any spectral line.\\

\begin{table*}
	\centering
	\caption{The table contains all the information regarding the rest wavelength estimation for each AR}
	\label{tab:rest_wave}
	\begin{tabular}{lcccccr}  
		\hline
    Sr. No. &  Type of AR & Cool spectral line & Centroid of cool spectral line (\AA) & Difference (Observed Centroid - Standard Centroid)(\AA) & Rest Wavelength for Si~{\sc iv} (\AA)\\
		\hline
  
		1 & Beta & Fe ~{\sc ii} & 1405.6050 & -0.0030 & 1402.7730\\
  \hline
		2 & Beta & S ~{\sc i} & 1401.5151 & 0.0016 & 1402.7716\\
  \hline
		3 & Beta & S~{\sc i} & 1401.5049 & -0.0086 & 1402.7786\\
  \hline
  	    4 & Beta & Fe ~{\sc ii} & 1405.5991 & -0.0090 & 1402.7790\\
  \hline
  	  5 & Beta & S~{\sc i} & 1401.5116 & -0.0019 & 1402.7719\\
  \hline
            6 & Beta-Gamma & Fe~{\sc ii} & 1405.5064 & -0.0072 & 1402.7772\\	  
  \hline
    	  7 & Beta-Gamma & S ~{\sc i} & 1401.5061 & -0.0074 & 1402.7774\\
  \hline
    	  8 & Beta-Gamma & Fe~{\sc ii} & 1405.6034 & -0.0066 & 1402.7766\\
  \hline
    	  9 & Beta-Gamma & Fe~{\sc ii} & 1405.6022 & -0.0078 & 1402.7778\\
  \hline
    	  10 & Beta-Gamma & S ~{\sc i} & 1401.5079 & -0.0056 & 1402.7756\\
  \hline
	\end{tabular}
\end{table*}
IRIS provides several cool lines that can be used in the rest wavelength estimation. And, we have utilized Fe~{\sc ii} and S~{\sc i} lines to calculate the rest wavelength. We have followed the same procedure as described in the IRIS Technical Note 20 available at IRIS/LMSAL website\footnote{\href{https://iris.lmsal.com/}{https://iris.lmsal.com/search/}}.
Finally, in table~\ref{tab:rest_wave}, we have tabulated all the necessary information related to the rest wavelength estimation, namely, used cool lines, observed centroid of cool lines, the difference between observed centroids and standard wavelength of cool lines (i.e., taken from CHIANTI database), and, the rest wavelength of Si~{\sc iv}. Using the estimated rest wavelength, the centroids are converted into the Doppler velocity.

\begin{figure*}
	\includegraphics[trim=1.0cm 0.2cm 2.0cm 0.0cm, scale=0.8]{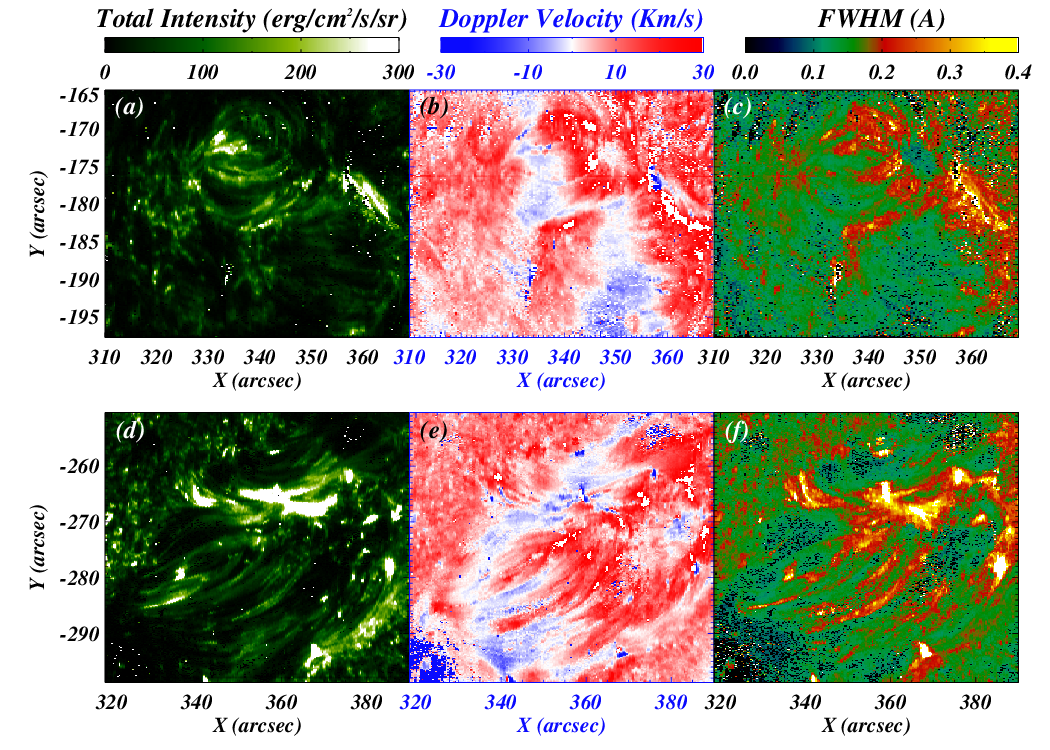}
    \caption{Top row shows the total intensity (panel a), Doppler velocity (panel b), and FWHM maps (panel c) of the first $\beta$ type AR. Similarly, the total intensity (panel d), Doppler velocity (panel e), and FWHM maps (panel f) of the first $\beta$-$\gamma$ type AR. We can clearly see the existence of various cool loops in the intensity maps of both types of ARs.} 
    \label{fig:fig2}
\end{figure*}

The integrated (total) calibrated intensity(panel a), Doppler velocity (panel b), and FWHM maps (panel c) of ROI from the first $\beta$ AR are displayed in the top row of figure~\ref{fig:fig2}. Similarly, the bottom row displays total calibrated intensity (panel d), Doppler velocity (panel e), and FWHM maps (panel f) of the first $\beta${--}$\gamma$ AR.\\
\section{Results}
\label{section:results}
\subsection{The Selection of the Foot Points of Cool loops}

\begin{figure*}
    \includegraphics[trim=4.0cm 0.2cm 3.0cm 0.0cm, scale=0.8]{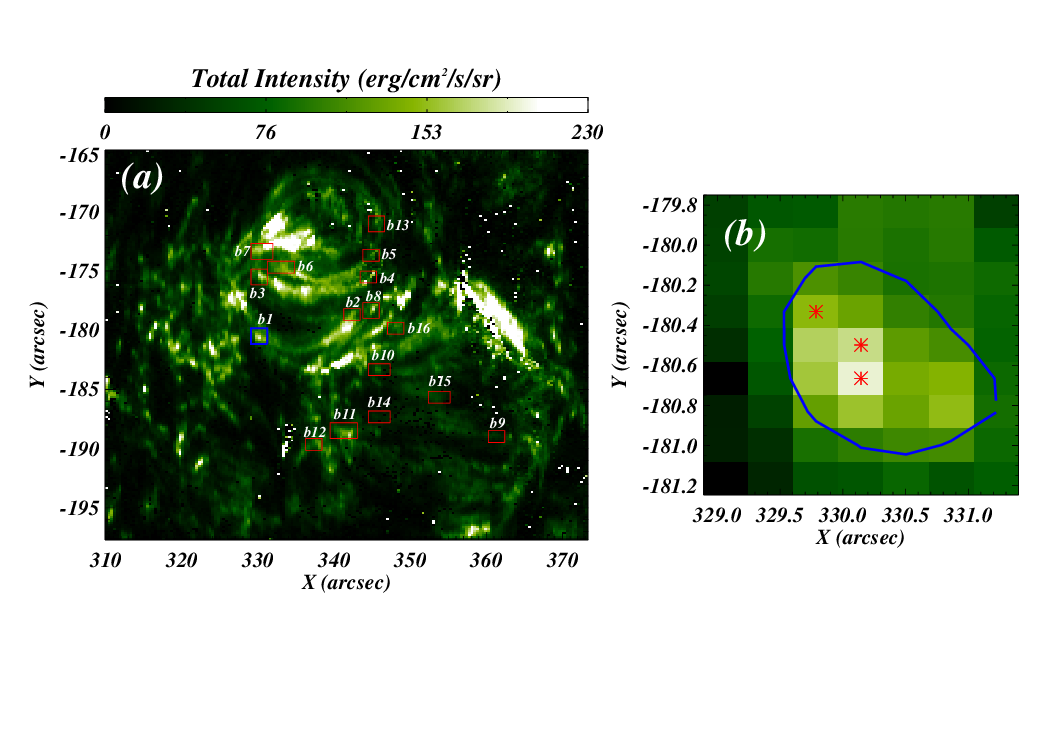}
    \\
    \includegraphics[trim=4.0cm 2.2cm 3.0cm 3.0cm, scale=0.8]{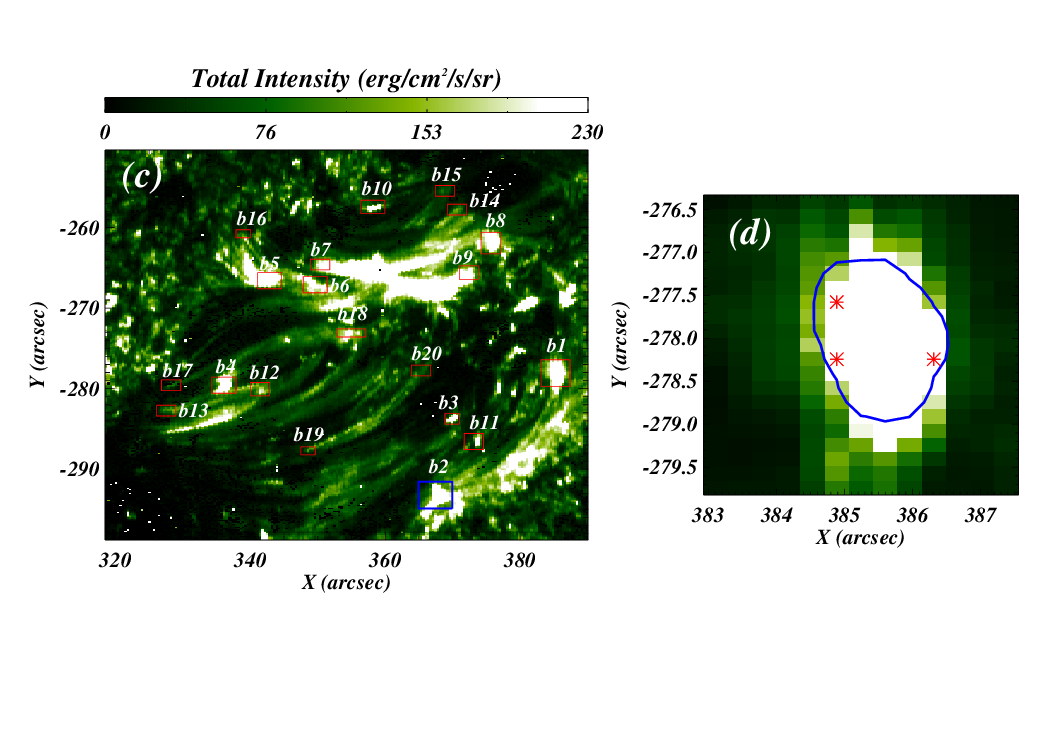}
   \caption{The panel (a) displays the total calibrated intensity map (ROI) of first $\beta$ type AR with 16 different boxes (shown by red rectangular box) at the cool loop footpoints. Please note that one particular box is shown in blue color in panel (a), and this box is displayed in panel (b). The blue contour outlines the footpoint region of the cool loops. The panels (c) and (d) are same as (a) and (b) but for the first $\beta$-$\gamma$ type AR with 20 different boxes at the cool loop footpoints}
    \label{fig:fig3}
\end{figure*}

We have displayed integrated calibrated intensity maps (ROI) of the first $\beta$ type AR panel (a) and the first $\beta$-$\gamma$ type AR panel (c) of the figure~\ref{fig:fig3}. Several footpoints of cool loops are clearly visible in both ARs (see panels (a) and (c)). We have selected, manually, 16 different boxes (displayed by red boxes) in the vicinity of the foot points of the cool loops in the $\beta$ type AR (panel a) and 20 boxes in the $\beta$-$\gamma$ type AR (panel c). One particular box, which is shown in blue color in panel (a), is further shown in panel (b) of figure~\ref{fig:fig3}. Similarly, the blue color box of panel (c) is displayed in panel (d) of figure~\ref{fig:fig3}.\\

Major area of both boxes shown in panels (b) $\&$ (d) has a bright region (i.e., foot point). However, a small fraction of the area of these boxes are less bright (i.e., non-footpoint region) and is most probably not related to the footpoints of the cool loops. To exclude this non-footpoint region in these boxes, we have calculated the mean intensity of that particular box (i.e., an average of the integrated calibrated intensities over all the pixels of that particular box. For instance, the average intensity of boxes shown in panels (b) and (d) are 94.82 and 751.34, respectively.) Further, using this mean intensity value as an intensity threshold, we draw the contour within each box, i.e., the blue contours in panels (b) and (d) of figure~\ref{fig:fig3}. Now, the region inside the blue contour is considered as the footpoint region while the region outside of the blue contour is the non-footpoint region. We have applied the above-described procedure to all boxes of all five $\beta$ and five $\beta${--}$\gamma$ type ARs. Here, it is to be noted that we have provided the mean calibrated intensity of each box from all $\beta$ and $\beta${--}$\gamma$ ARs in the appendix~\ref{sect:appendix_mean_intensity}.\\

\begin{figure*}
	\includegraphics[trim=3.0cm 0.2cm 2.0cm 0.0cm, scale=0.8]{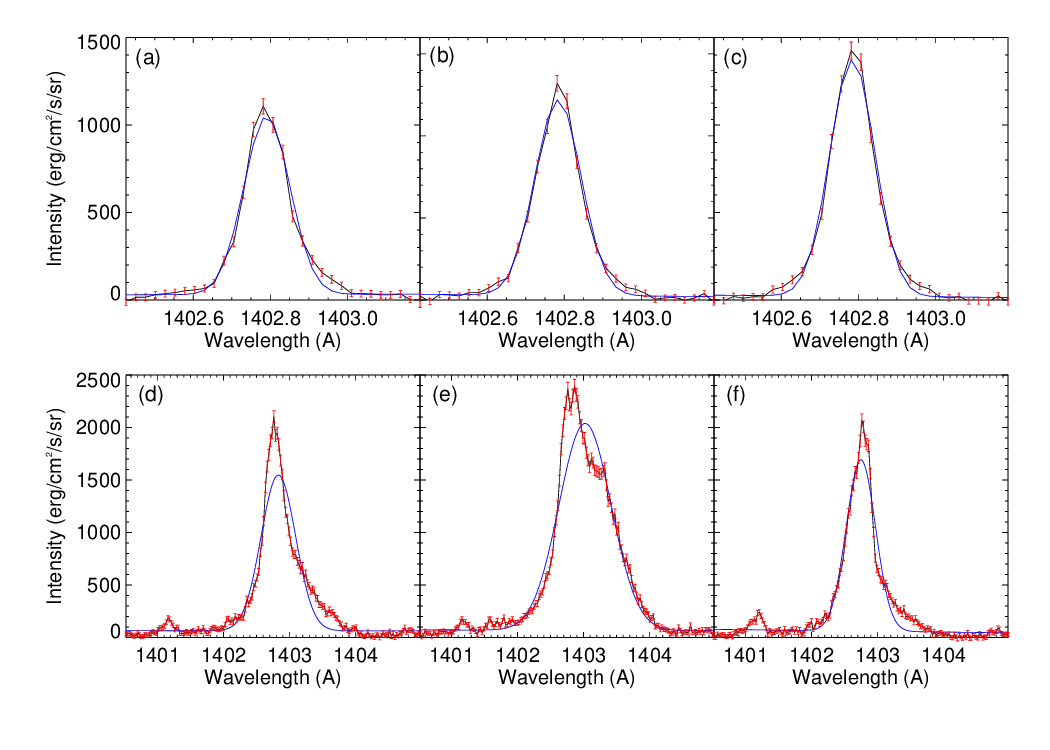}
    \caption{All three panels of the top row displays the spectral profiles of three different locaations marked by the asterisk sign in the panel (b) figure~\ref{fig:fig3}. Similarly, the bottom row shows the spectral profiles from three locations that are marked by the asterisk in the panel (d) figure~\ref{fig:fig3}. The overplotted blue curve in all panels are the single Gaussian fit on the observed spectral profiles.} 
    \label{fig:fig4}
\end{figure*}

Then, we collected integrated calibrated intensity, integrated uncalibrated intensity Doppler velocity, and FWHM from each location of the footpoints, i.e., from only bright regions in each box of all five $\beta$ type ARs and $\beta${--}$\gamma$ type ARs. To show the spectra from loop footpoints, we have chosen three locations in the vicinity of footpoints of $\beta$ AR (see asterisk signs in panel b; Figure~\ref{fig:fig3}) and three more locations in $\beta$-$\gamma$ AR (see, three asterisk signs in panel (d); Figure~\ref{fig:fig3}). The top row of Figure~\ref{fig:fig4} shows the spectra (black solid curve) from the three locations of the panel (b) of Figure~\ref{fig:fig3}. Similarly, spectra from the three locations of footpoints of $\beta$-$\gamma$ AR are displayed in the bottom row of Figure~\ref{fig:fig4}. The blue curve in all panels of Figure~\ref{fig:fig4} are the Gaussian fit on the corresponding observed profile. Here, it is clearly visible that spectral profiles in $\beta$-$\gamma$ AR are far more complex than the spectral profiles in $\beta$ AR.

\subsection{Statistical Analysis of the Footpoints of Cool loops}

\begin{figure*}
    \includegraphics[trim=2.0cm 0.0cm 3.0cm 0.0cm, scale=1.0]{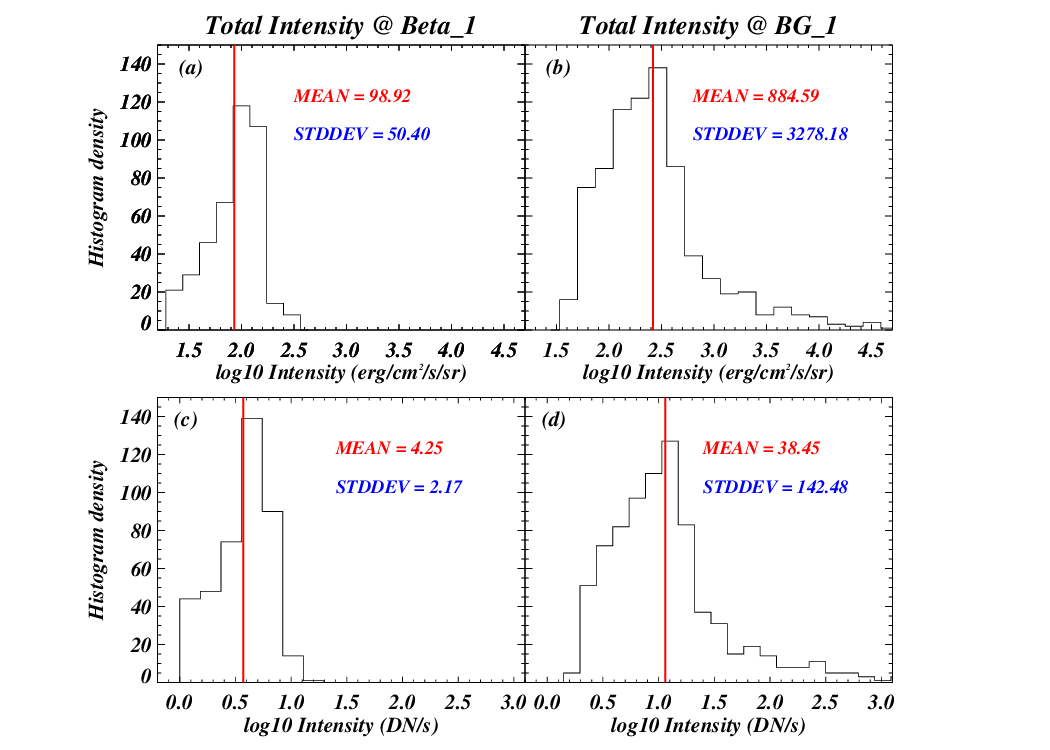}
   \caption{The calibrated intensity histograms of the first $\beta$ and first $\beta${--}$\gamma$ ARs. The mean intensities are 98.92 (erg/cm$^{2}$/s/sr) and 884.59 (erg/cm$^{2}$/s/sr) in first $\beta$ (panel (a)) and first $\beta${--}$\gamma$ ARs (panel (b)). The panels (c) and (d) show the uncalibrated intensity histograms, and the mean uncalibrated intensities are 4.25 (DN/s) and 38.45 (DN/s) in the first $\beta$ and first $\beta${--}$\gamma$, respectively.}
    \label{fig:fig5}
\end{figure*}

First of all, we would like to mention that the ARs, which we have used in the present analysis, span the period of more than 9 years, i.e., from 2013 to 2023 (see table~\ref{tab:obs_details}). We know that all the instruments, including IRIS, degrade with time. Hence, the instrumental degradation aspect should be included in this comparative analysis. Therefore, we have compared the calibrated as well as uncalibrated intensities of the loop footpoints (Figure~\ref{fig:fig5}) from two ARs that are close in time, i.e., first $\beta$ AR (i.e., AR11937) that was observed on 2014-01-01 and the first $\beta$-$\gamma$ AR (i.e., AR11934) that was observed on 2013-12-27. The time difference between AR11937 and AR11934 is just 4 days. Hence, the instrument degradation effect would have not been present (at least) between these two ARs.\\ 

The mean of integrated calibrated (erg/cm$^{2}$/s/sr) intensity of the loop footpoints is $\sim$99 and 885 in the first $\beta$ (panel a; Figure~\ref{fig:fig5}) and first $\beta${--}$\gamma$ ARs (panel b; Figure~\ref{fig:fig5}), respectively. It means that the integrated calibrated intensity of the loop's footpoints in the first $\beta${--}$\gamma$ AR is nearly 9 times the corresponding integrated calibrated intensity in the $\beta$ AR. Similarly, the mean of uncalibrated intensity (i.e., DN/s) of loop footpoints in the first $\beta${--}$\gamma$ AR (i.e., 38.45; panel d of Figure~\ref{fig:fig5}) is also nearly 9 times the corresponding mean of uncalibrated intensity of first $\beta$ AR (i.e., 4.25; panel c; Figure~\ref{fig:fig5}). Hence, we can say that the calibrated as well as uncalibrated intensities of the loop footpoints do show almost the same behavior.\\

\begin{figure*}
    \includegraphics[trim=4.0cm 0.0cm 4.0cm 0.0cm, scale=1.0]{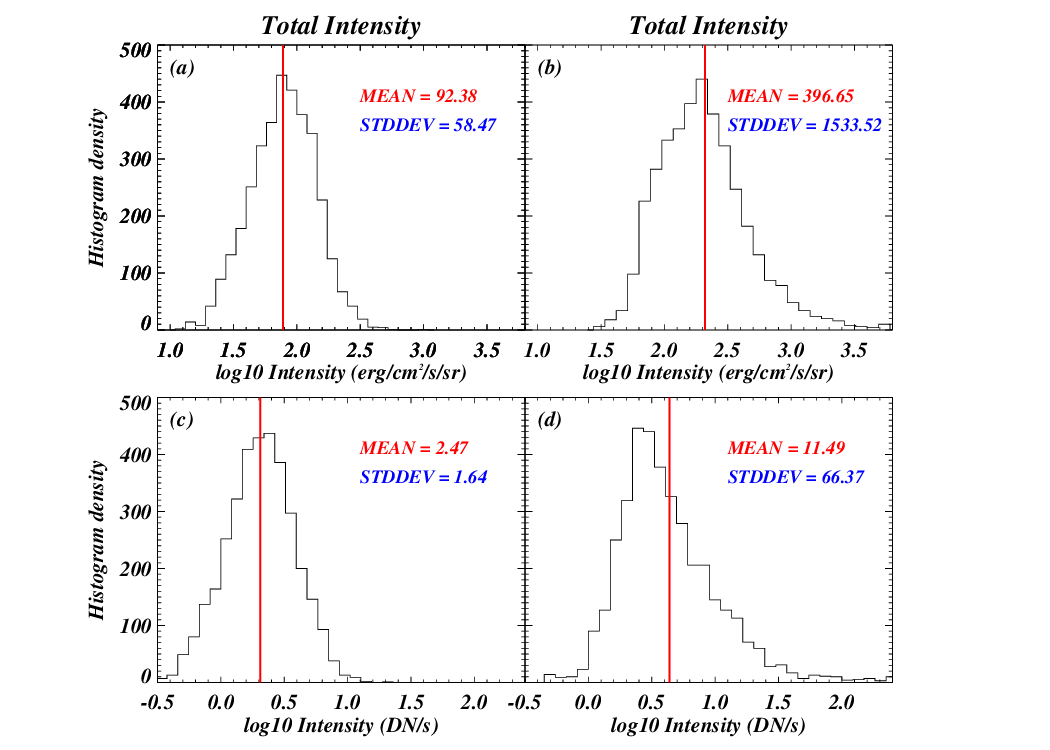}
    \caption{Same as Figure~\ref{fig:fig5} but from all five $\beta$ and five $\beta${--}$\gamma$ ARs.}
    \label{fig:fig6}
\end{figure*}

Further, in the next step, we have produced the histograms of the calibrated intensities and uncalibrated intensities of the loop footpoints from all five $\beta$ and $\beta$-$\gamma$ ARs (Figure~\ref{fig:fig6}). Now, the mean calibrated intensity of the loop footpoints in $\beta${--}$\gamma$ type ARs (panel b) is more than 4 times the intensity of the loop footpoints in $\beta$ type ARs (panel a; Fig~\ref{fig:fig6}). The mean uncalibrated intensity of the loop footpoints in $\beta${--}$\gamma$ ARs (panel d) is also more than 4 times the intensity of the loop footpoints in $\beta$ ARs (panel c; Fig~\ref{fig:fig5}). Hence, the same intensity difference is reduced to less than half (i.e., 4 times) if we consider all the ARs together (over different years), see Figure~\ref{fig:fig6}.

Most of the $\beta$ ARs (i.e., four out of five observations)  were observed from January 2014 to December 2017. Most of the $\beta${--}$\gamma$ ARs (four out of five) were observed from June 2018 to February 2023. Hence, the effect of instrumental degradation is more on the $\beta${--}$\gamma$ ARs than $\beta$ ARs. Even after the dominant effect of instrumental degradation on $\beta${--}$\gamma$ ARs, the mean intensity of $\beta${--}$\gamma$ ARs is more than 4 times higher than the corresponding intensity of $\beta$ ARs. Although, the same difference is found to be around 9 times for both type ARs observed nearly at the same time (Figure~\ref{fig:fig5}). The reduction in the intensity difference (when we consider all the ARs together) is due to the instrumental degradation as the ARs cover a period of more than 9 years. \\

Perhaps, the footpoints of cool loops may be contaminated by the emission coming from the lower solar atmosphere, i.e., plage emission from the solar chromosphere. In the current observational baseline, it is hard to estimate the plage contribution into the loop footpoints. However, we have tried to understand the contribution by the plage into the loop footpoints. For this, we have analyzed chromospheric intensity of all boxes in the first $\beta$ and first $\beta${--}$\gamma$ ARs (as shown in Figure~\ref{fig:fig3}) using IRIS/SJI 2796~{\AA}. We have found that mean chromospheric intensity (i.e., averaged over all boxes) is nearly same in both types of ARs. Hence, if plage emission (i.e., chromospheric emission) contributes to the loop footpoints in TR then, this contribution is almost same in the footpoints of cool loops of $\beta$ and $\beta${--}$\gamma$ ARs. Therefore, we are lead to believe that differences in the loop footpoint intensities are due to the TR activities, and not due to the plage contamination.





\begin{figure*}
    \includegraphics[trim=3.0cm 0.2cm 3.0cm 0.0cm, scale=0.9]{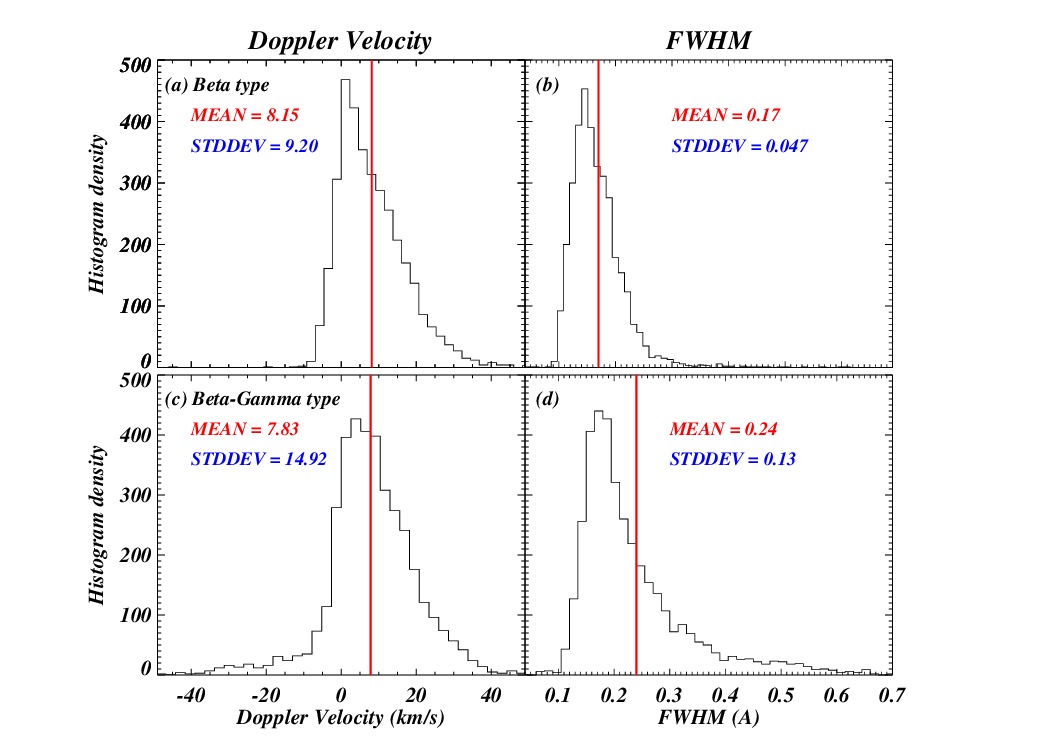}
    \caption{The panels (a) and (c) show the Doppler velocity histograms of $\beta$ and $\beta${--}$\gamma$ ARs. The mean Doppler velocities are 8.15$\pm$9.20 km/s and 7.83$\pm$14.92 km/s in $\beta$ and $\beta${--}$\gamma$ ARs, respectively. Similarly, the panels (b) and (d) are dedicated for FWHM in $\beta$ and $\beta${--}$\gamma$ ARs. The mean FWHM in $\beta$ ARs is 0.17$\pm$0.047~\AA while it is 0.24$\pm$0.13~\AA in $\beta${--}$\gamma$ ARs. The vertical red solid lines in all the panels represent the corresponding mean values.}
    \label{fig:fig7}
\end{figure*}

The histograms of the Doppler velocity and FWHM of all five $\beta$ and five $\beta${--}$\gamma$ ARs are shown in figure~\ref{fig:fig7}. The mean Doppler velocity in $\beta$ ARs is 8.15$\pm$9.20 km/s (panel a) and in $\beta${--}$\gamma$ ARs is 7.83$\pm$14.92 km/s (panel c). Then, the mean FWHM of $\beta$ and $\beta${--}$\gamma$ ARs are 0.17$\pm$0.047~{\AA} (panel b) and 0.24$\pm$0.13~{\AA} (panel d), respectively. The solid red vertical line in each panel of figure~\ref{fig:fig7} shows the mean values of the corresponding parameters. Further, we have mentioned mean and standard deviation values in each panel. The mean FWHM of the loop footpoints in $\beta$-$\gamma$ ARs is significantly higher (i.e., almost 1.5 times) than the mean FWHM of $\beta$ ARs. Here, it should also be noted that the spread of the FWHM histograms of $\beta${--}$\gamma$ ARs is around 3 times higher than the corresponding spread of FWHMs in $\beta$ ARs. This higher FWHM and more spread justify the higher activity level at the footpoints of cool loops in $\beta$-$\gamma$ ARs. On the contrary the mean Doppler velocity of both $\beta$ and $\beta${--}$\gamma$ ARs are almost the same. But the spread of the Doppler velocity in $\beta$-$\gamma$ ARs is more than 1.5 times higher than the corresponding spread of Doppler velocity in $\beta$ ARs. On the contrary, the mean Doppler velocities for both types of ARs are almost the same (see panels a and c of Figure~\ref{fig:fig7}). Here, it should be noted that the spread of Doppler velocity distribution in $\beta$-$\gamma$ ARs is more than 1.5 times the corresponding spread of Doppler velocity in $\beta$ ARs.\\

\begin{figure*}
    \includegraphics[trim=3.0cm 0.0cm 3.0cm 0.0cm, scale=1.1]{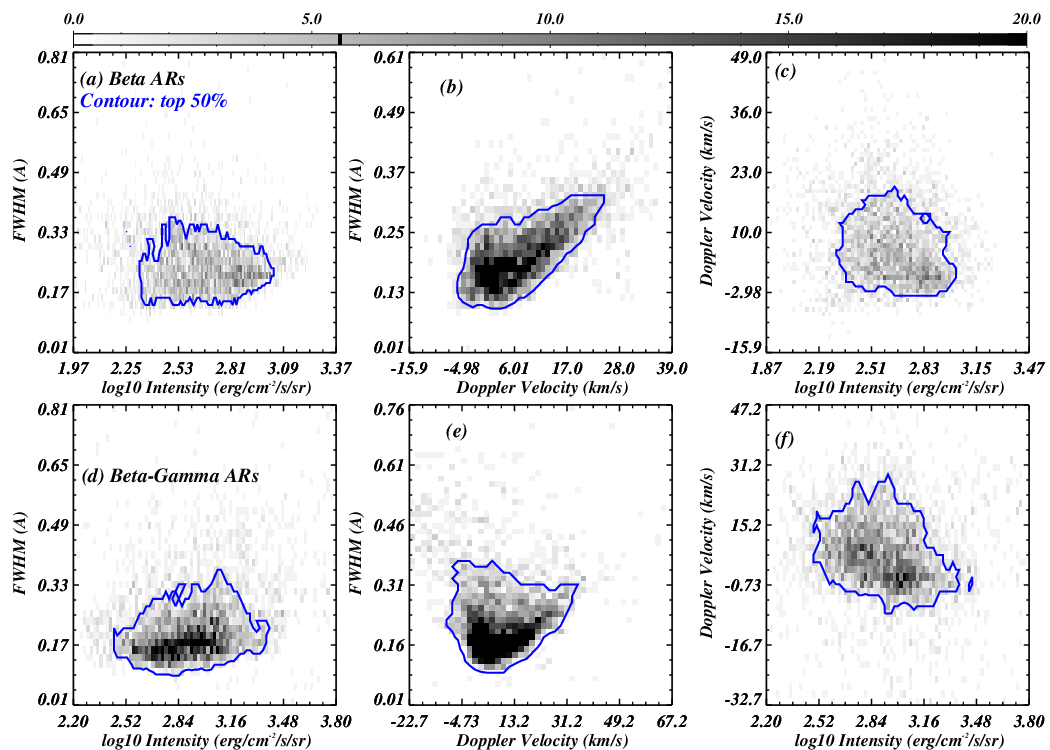}
    \caption{The figure shows 2-D histograms of various correlations among the peak calibrated intensity, Doppler velocity, and FWHM of the loops footpoints, namely, peak calibrated intensity vs FWHMs for $\beta$ (panel a) and $\beta$-$\gamma$ type ARs (panel d), Doppler velocity vs FWHMs for $\beta$ (panel b) and $\beta$-$\gamma$ type ARs (panel e), and Doppler velocity vs peak calibrated intensity for $\beta$ (panel c) and $\beta$-$\gamma$ type ARs (panel f). In each 2-D histogram (i.e., each panel), the blue contours outline the top 50\% of the histogram frequency. The black in all 2D histograms indicates the highest frequency, and all these 2D histograms are plotted on a linear scale.}
    \label{fig:fig8}
\end{figure*}
The 2D histograms of peak calibrated intensity{--}FWHM (panel a), Doppler Velocity{--}FWHM (panel b), and peak calibrated intensity{--}Doppler Velocity (panel c) for the footpoints of the loops of $\beta$ type ARs are displayed in the figure~\ref{fig:fig8}. The same correlations for the loop footpoints in $\beta${--}$\gamma$ ARs are displayed in the panels (d), (e), and (f) of figure~\ref{fig:fig8}. The white color corresponds to the lowest histogram density (i.e., zero density) while the black color shows the highest histogram density in all 2D histograms. The colorbar at the top of this figure~\ref{fig:fig8} represents the variations of histogram density. We sorted the non-zero densities of each histogram, and then we estimated the histogram density value at the middle location. For example, 2D histograms of peak calibrated intensity{--}FWHM (panel a) have a total of 1286 non-zero histogram density points. Now, we sorted 1286 histogram density points in increasing order, and we found that the histogram density value at the middle location (i.e., at location number 643) is 5. Then, we used this value (i.e., 5) as a threshold value to draw contour on the 2D histograms of peak calibrated intensity{--}FWHM (see blue contour; panel a). The same procedure is applied to other 2D histograms to draw the blue contours. Finally, we mention that blue contours in each 2D histogram enclose the highest 50 \% histogram densities.\\

In the case of $\beta$ ARs, the peak calibrated intensities and FWHM are not correlated (panel a). However, in the case of $\beta$-$\gamma$ ARs, peak calibrated intensity are weakly (positively) correlated with FWHM (panel d). The Doppler velocity is well (positively) correlated with FWHM in both types of ARs (see panels (b) and (e)). While, on the contrary, we do not see any profound correlation between peak calibrated intensity and Doppler velocity of the footpoints of the cool loops in the $\beta$ type ARs (panel c) and  $\beta$-$\gamma$ type ARs (panel f).

\subsection{Line Ratios of Si~{\sc iv} ($\lambda$1394/$\lambda$1403) in the Footpoints of loops}
\begin{figure*}
    \includegraphics[trim=4.0cm 3.0cm 4.0cm 1.0cm, scale=1.0]{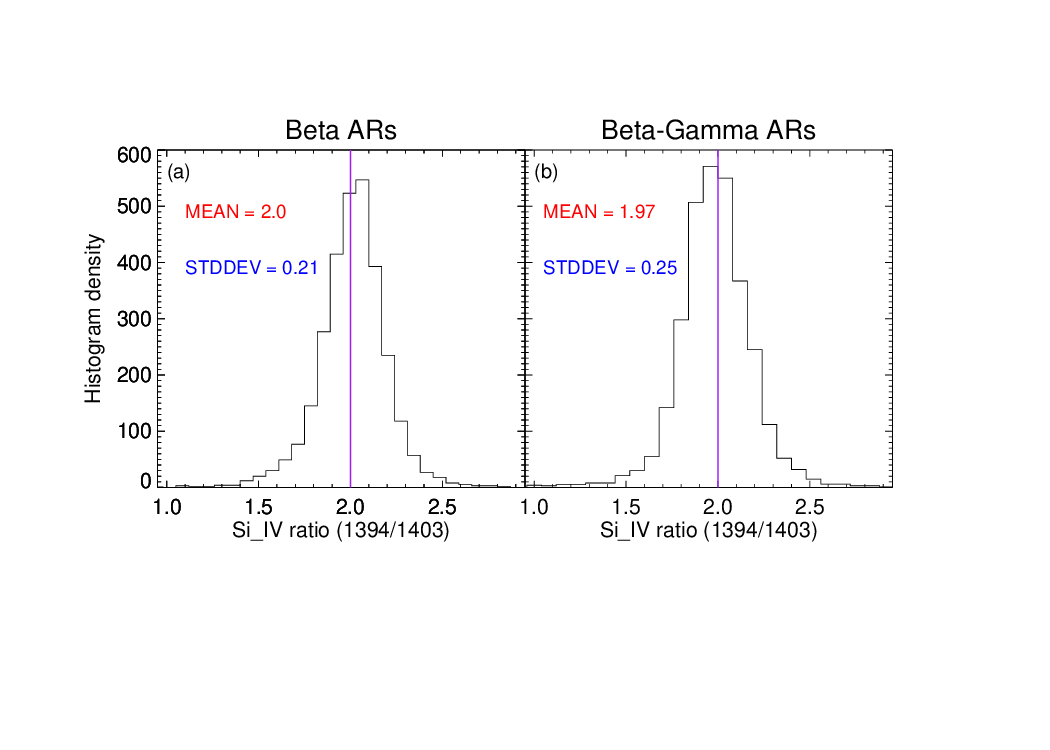}
    \caption{The histograms of Si~{\sc iv} line ratio from the loop footpoints of $\beta$ (panel a) and $\beta$-$\gamma$ type ARs (panel b). The vertical purple lines in both panels are located at the line ratio of 2, which is the theoretical line ratio of these two Si~{\sc iv} lines. The mean ratio value in $\beta$ ARs is exactly equal to the theoretical value while the mean value in $\beta${--}$\gamma$ ARs is slightly less than the theoretical value, i.e., 1.97.}
    \label{fig:fig9}
\end{figure*}
The intensity ratio of Si~{\sc iv} resonance lines (i.e., $\lambda$1393.75/$\lambda$1403.77) is an important measure of the opacity. Generally, it is assumed that Si~{\sc iv} lines form in optically thin conditions. However, this is not the case always, and Si~{\sc iv} spectral lines may also form in the optically thick conditions (e.g., \citealt{Yan2015, DT2020, Kayshap2021}). Theoretically, we know that the intensity of Si~{\sc iv} 1393.755~{\AA} is double the intensity of  Si~{\sc iv} 1402.77~{\AA} (e.g., \citealt{Dere1997, Dere2019}). Hence, the intensity ratio must be 2 if the lines are forming in optically thin conditions. Hence, simply, with the help of this intensity ratio, we can understand whether the line is forming under optically thin or thick conditions. (\citealt{DT2020, Kayshap2021}).\\    

The figure~\ref{fig:fig9} shows the histograms of intensity ratio for the loop footpoints in $\beta$ ARs (left panel) as well as $\beta${--}$\gamma$ ARs (right panel). In both histograms, the vertical solid purple line shows the location of the theoretical intensity ratio (i.e., 2.0). The ratio values are almost ranging from 1.3 to 2.6 in both AR types. The mean ratio values are 2.0 and 1.97 for $\beta$ and $\beta${--}$\gamma$ ARs. Hence, the mean ratio is found to be slightly low for the footpoints in $\beta$-$\gamma$ type ARs than that for $\beta$ type ARs. The spread of ratio values (i.e., standard deviation) is more in $\beta$-$\gamma$ type ARs (0.25) than $\beta$ type ARs (i.e., 0.21).\\

Further, we have searched the locations where the ratio significantly deviates from the theoretical ratio (i.e., 2). To know this, first of all, we have chosen the locations where the line ratios are less than 1.90 and the line ratios are greater than 2.10. In the case of the footpoints of $\beta$ ARs, there are 23\% locations which are having a ratio of less than 1.90, and almost 29\% locations which are having a ratio greater than 2.10. Similarly, in the case of the footpoints of $\beta$-$\gamma$ ARs, 31\% locations are having a ratio less than 1.90, and almost 24\% locations are having ratios greater than 2.10. In total, around 52\% locations in $\beta$ ARs and 55\% locations in $\beta$-$\gamma$ ARs are significantly deviated from the theoretical ratio. Hence, it is factually established that Si~{\sc iv} spectral lines originating from more than half of the locations of the footpoints of cool loops are formed in optically thick conditions.\\

\begin{figure*}
    \includegraphics[trim=4.0cm 1.0cm 4.0cm 0.8cm, scale=1.0]{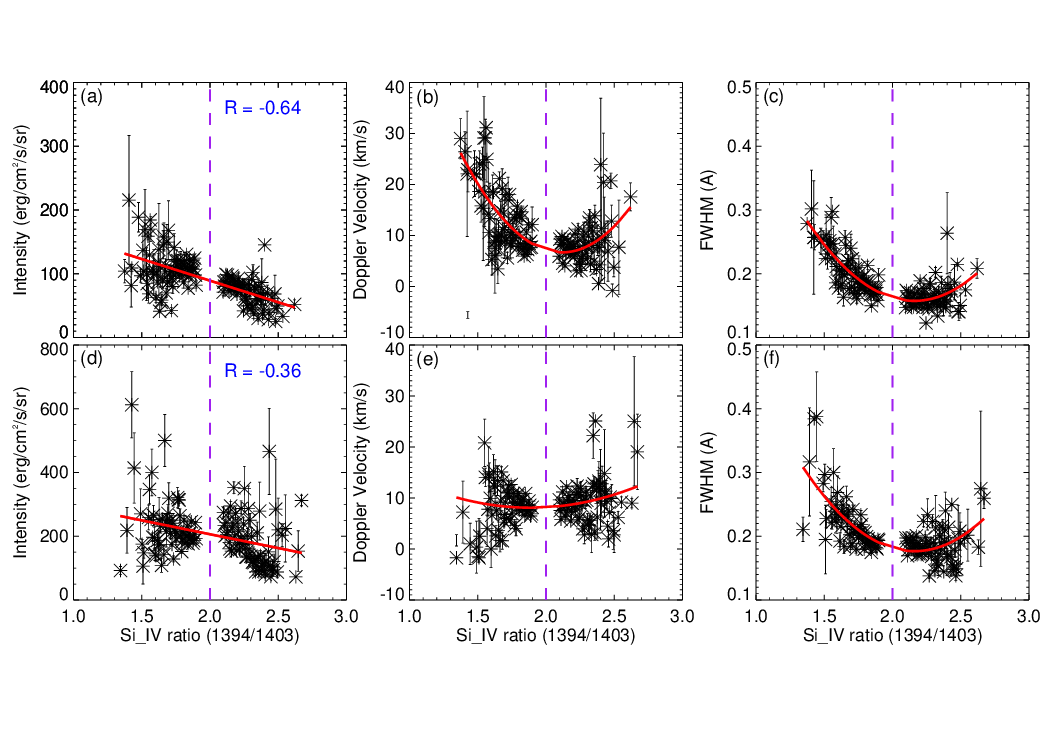}
    \caption{The figure displays some important correlation of line ratio, which is significantly deviated from theoretical ratio, with spectroscopic parameters of footpoints, namely, line ratio vs intensity for $\beta$ (panel a) and $\beta${--}$\gamma$ ARs (panel d), line ratio vs Doppler velocity for $\beta$ (panel b) and $\beta${--}$\gamma$ ARs (panel e), and line ratio vs FWHM for $\beta$ (panel c) and $\beta${--}$\gamma$ ARs (panel f). The intensity becomes stronger as the line ratio drops below 1.90 while the intensity decreases when the line ratio increases beyond 2.10 (see panels a and d). The Doppler velocity as well as FWHM increase in both cases, i.e., either the line ratio drops below 1.90 or increases beyond 2.10.}
    \label{fig:fig10}
\end{figure*}

We have created 500 bins from the significant line ratios (i.e., only the line ratios less than 1.90 and greater than 2.10). Further, we have collected the corresponding parameters (i.e., intensity, Doppler velocity, and FWHM) in each ratio bin, and then calculated the mean and standard error of each parameter in each ratio bin. Hence, through this approach, we have obtained 500 mean values and standard errors for each parameter. Then, using these mean values and standard errors, we have performed some correlations shown in figure~\ref{fig:fig10}, namely, (1) line ratio vs. total calibrated intensity for $\beta$ ARs (panel a) and $\beta$-$\gamma$ ARs (panel d), (2) line ratio vs. Doppler velocity for $\beta$ ARs (panel b) and $\beta$-$\gamma$ ARs (panel e), and (3) line ratio vs. FWHM for $\beta$ ARs (panel c) and $\beta$-$\gamma$ ARs (panel f). In both $\beta$ and $\beta${--}$\gamma$ ARs, the total calibrated intensity is high when the Si~{\sc iv} ratio is less than 1.90. If the Si~{\sc iv} ratio is greater than 2.10 then the total calibrated intensity is low (see panels (a) and (d)). On the contrary, for $\beta$ and $\beta${--}$\gamma$ ARs, the redshifts (Doppler velocity) increase in both cases whether the line ratio goes down below 1.90 or it increases beyond 2.10 (see panels b and e). The same behavior is found for FWHM in both types of ARs, see panels (c) and (f). Here, it is to be noted that the above-described pattern of Doppler velocity with line ratio is weak in $\beta${--}$\gamma$ ARs (panel e) in comparison to the same correlation for $\beta$ ARs (panel b). However, the correlation of FWHM with line ratio is strong in both types of ARs (see panels c and f).

\subsection{Complexity of Spectral Profiles}
\begin{figure*}
    \includegraphics[trim=4.0cm 3.0cm 3.0cm 1.0cm, scale=0.9]{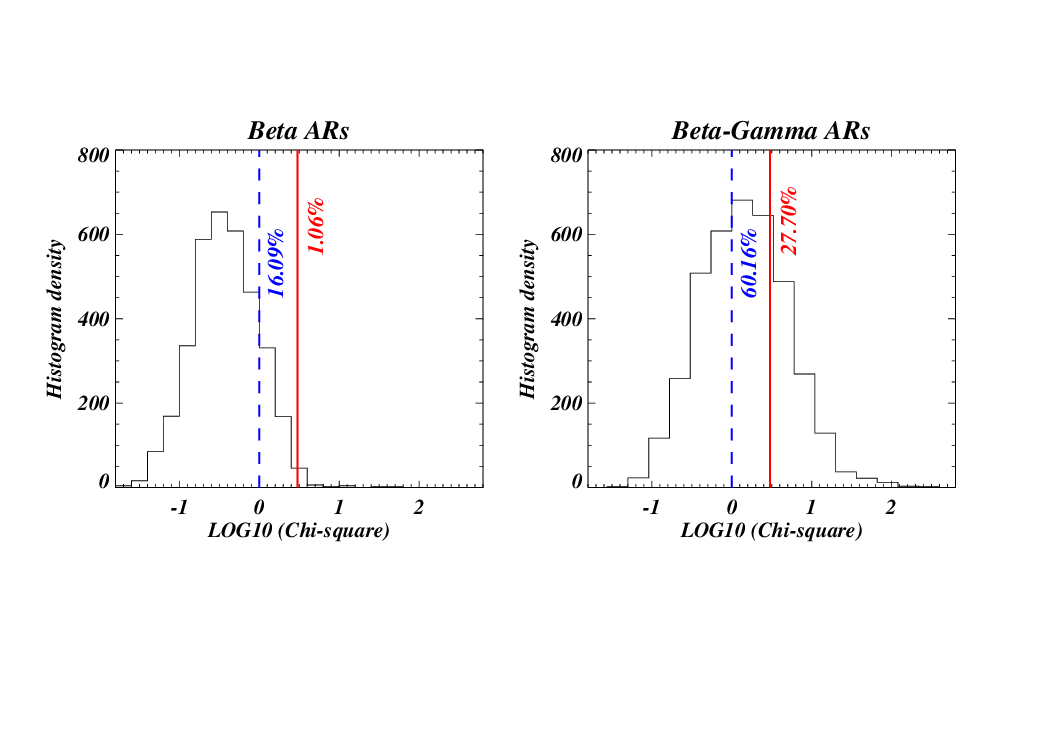}
    \caption{The histogram of chi-square values deduced from each location of loop footpoints from $\beta$ (left-panel) and $\beta$-$\gamma$ ARs (right-panel). The vertical blue dashed line and red solid line in both panels are located at the reduced chi-square value of 1 and 3, respectively. In the present analysis, the profiles that have a chi-square value greater than 3 is considered complex profile. The complex profiles are just 1.06\% in $\beta$ type ARs while the complex profiles are more than 26 times higher (i.e., 27.70\%) in the $\beta$-$\gamma$ type ARs than $\beta$ type ARs.}
    \label{fig:fig11}
\end{figure*}

Finally, during the analysis, we noticed that not all the spectral profiles are single Gaussian, but some spectral profiles deviate from the single Gaussian. We have estimated the chi-square value for each location in the footpoints of cool loops, and we produced the histogram of $\beta$ (left-panel; Figure~\ref{fig:fig7}) as well as $\beta$-$\gamma$ type ARs (right-panel; Figure~\ref{fig:fig11}). 
The blue dashed vertical line is located at the reduced chi-square value of 1 in both panels.  Interestingly, there are more than 16\% complex profiles (i.e., profiles having reduced chi-square values greater than 1) in the loop footpoints of $\beta$ ARs. While $\beta$-$\gamma$ ARs have more than 60\% complex profiles. Hence, we can say that the footpoints of the loops in $\beta$-$\gamma$ type ARs have $\sim$3.7 times more complex profiles than that in $\beta$ type ARs.\\

In general, if the chi-square value is greater than one, the profile is said to be asymmetric. However, we can consider the profile as a single Gaussian up to a specific chi-square value, for example, \citealt{YR2022} shows that a spectral profile with a chi-square value less than 5 can be effectively fitted with a single Gaussian. However, they drew their conclusion based on QS observations. While we are dealing with ARs, and, therefore, the chi-square threshold value may differ.\\
We have investigated the spectral profiles with respect to the chi-square value (see appendix~\ref{sect:appendix}), and it is found that the spectral profiles having chi-square values less than 3 can be well fitted with a single Gaussian (appendix~\ref{sect:appendix})). Therefore, in the present work, we used 3 as a chi-square threshold value to define the complex profiles. The solid red lines, in both panels, are at the threshold value of the chi-square (i.e., 3). Interestingly, the very complex profiles in the loop footpoints of $\beta${--}$\gamma$ ARs are more than 26 times the complex profiles in $\beta$ type ARs. This observational finding justifies that the footpoints of cool loops have high complex behaviour in $\beta$-$\gamma$ type ARs than the simple $\beta$ type ARs.\\ 
\section{Summary and Discussion}
The present work is to statistically investigate the footpoints of cool loops using IRIS and HMI/SDO observations. IRIS has revealed that cool loops are an integral part of ARs (e.g., \citealt{Huang2015, Rao2019a, Rao2019b}). In the present work, we have taken five $\beta$ type ARs and five $\beta${--}$\gamma$ type ARs to diagnose the footpoint of cool loops. Further, a comparison is also performed between the properties of the footpoints of these cool loops in $\beta$ and $\beta${--}$\gamma$ type ARs. Through the statistical approach, we have derived some important findings about the footpoints of cool loops, which are summarized below.

\begin{itemize}

   \item The mean calibrated and uncalibrated intensities from first $\beta$ AR are 98.92$\pm$50.40 erg/cm$^{2}$/s/sr and 4.25$\pm$2.17 DN/s, respectively. The same intensities for first $\beta${--}$\gamma$ AR are 884.59$\pm$3278.18 erg/cm$^{2}$/s/sr and 38.45$\pm$142.48 DN/s.

    \item Similarly, the mean calibrated and uncalibrated intensities from all five $\beta$ ARs are 92.38$\pm$58.47 erg/cm$^{2}$/s/sr and 2.47$\pm$1.64 DN/s, respectively. The same intensities are 396.65$\pm$1533.52 erg/cm$^{2}$/s/sr and 11.49$\pm$66.37 DN/s but for all five $\beta${--}$\gamma$ ARs. 
    
    \item The mean Doppler velocity and FWHM of the footpoints of the cool loops in all five $\beta$ type ARs are 8.15$\pm$9.20 km/s, and 0.17$\pm$0.047~{\AA}, respectively. The same parameters are 7.83$\pm$14.92 km/s and 0.24$\pm$0.13~{\AA} but for all five $\beta${--}$\gamma$ ARs.  

    \item The mean calibrated and uncalibrated intensities are 9 times stronger in the $\beta${--}$\gamma$ AR in comparison to the $\beta$ AR if we consider both AR close in time, i.e., no instrumental degradation. The same intensity difference is reduced to 4 times if we consider all the ARs. These ARs are observed over 9 years, therefore, the reduction in the intensity difference is most probably due to the instrumental degradation.
    
    \item There is no correlation between the central (peak) calibrated intensity and FWHM, and between the central calibrated intensity and Doppler velocity of the footpoints of the loops in $\beta$ as well as $\beta$-$\gamma$ type ARs. However, the Doppler velocity and FWHM of the cool loop footpoints in both types of ARs are weakly correlated. 

     \item The Si~{\sc iv} line ratios from the footpoints of cool loops in both types of ARs are significantly deviated from the theoretical intensity ratio of Si~{\sc iv}. Here, we find that the opacity is important for more than half of the footpoint locations. 

     \item If the line ratio drops below 1.90 then the intensity of the footpoints becomes stronger. The intensity is getting weaker when the ratio goes beyond 2.10. However, the Doppler velocity and FWHM increase for low (i.e., the ratio less than 1.90) as well as high ratio values (i.e. ratio greater than 2.10).
    
     \item Lastly, we find that the footpoints of cool loops have more complex profiles in $\beta$-$\gamma$ type ARs (i.e., 27.70\%) than that in $\beta$ type ARs (i.e., 1.06\%). The complex profiles are almost 26 times higher in $\beta$-$\gamma$ type ARs than that in the $\beta$ type ARs. 
\end{itemize}

This is the first-ever research that demonstrates the complexity of ARs affecting, directly, the mean attributes of the footpoints of the cool loops. We examined a total of 200 footpoints of the cool loops from both types of ARs, i.e., 120 footpoints in $\beta$ type ARs and 80 footpoints in $\beta${--}$\gamma$ type ARs. Statistically, we have found significantly higher mean intensities of the loop footpoints in the complex $\beta${--}$\gamma$ type ARs than that in $\beta$ type ARs. We know that any instrument including IRIS degrades with time. Hence, the intensities from $\beta${--}$\gamma$ are significantly affected due to the instrumental degradation because most of $\beta${--}$\gamma$ ARs are observed later than $\beta$ ARs. Hence, the intensity differences between the loop footpoints of $\beta${--}$\gamma$ ARs and $\beta$ ARs (, i.e., loop footpoints in $\beta${--}$\gamma$ ARs have 4 times higher intensity than the corresponding intensity in $\beta$ ARs) are underestimated.\\

The plasma flows within the footpoints of cool loops are almost similar in both types of AR. In this analysis, we have considered blueshifted and redshifted footpoints of cool loops, and it is found that the majority of the footpoints are redshifted. The mean redshifts of these footpoints are around 8.0 km/s in both types of AR. In general, solar TR of quiet-Sun, coronal hole, and ARs are red-shifted (\citealt{Teriaca1999, Peter1999, PJ1999, Dadashi2011, Kayshap2015, Kayshap2017}). In the present analysis, similar to previous works, we have also found that the footpoints of the cool loop in both ARs are redshifted. Please note that the Doppler velocity of ARs varies as per the location of the AR on the solar disk, i.e., center-to-limb variations (CLV) do exist in the Doppler velocity of ARs (e.g., \citealt{Ghosh2019, Rajhans2023}). The Doppler velocity is maximum at the disk center which decreases while moving toward the solar limb. The Doppler velocity near the disk center is around 7{--}9 km/s (\citealt{Rajhans2023}). In the present work, most of the ARs are close to the disk center (see the $\mu$ values in table~\ref{tab:obs_details}), and the on-average Doppler velocity of footpoints of cool loops is $\approx$ 8.0 km/s. It (i.e., the mean value of Doppler velocity) is consistent with previously reported Doppler velocity of the footpoints of cool loops (e.g., \citealt{Rao2019b, Rao2019a}) and ARs (\citealt{Rajhans2023}).\\

The footpoints of the cool loops exhibit large variations in the FWHM as shown by \cite{Rao2019b}, i.e., 0.095{--}0.246 (first set of observations), 0.094{--}0.200 (second set of observations), and 0.125{--}0.182 (third set of observations). Earlier, \cite{Huang2015} also showed broad spectral profiles at the footpoints of cool loops. Recently, \cite{Srivastava2020} have found large FWHM (more than 0.25~{\AA}) around the footpoint of the cool loops. The present work systematically estimates the FWHM in the footpoints of cool loops in both types of ARs. The spreads of FWHM are 0.10{--}0.40~{\AA} and  0.10{--}0.70~{\AA} in the footpoints of the cool loops in $\beta$ and $\beta$ $\gamma$ ARs, respectively, and the reported range is consistent with the previous findings (e.g., \citealt{Rao2019a, Rao2019b}). Most importantly, the mean FWHM in the loop footpoints is higher in $\beta${--}$\gamma$ ARs than $\beta$ ARs. In addition, the spread in FWHM is more than 2 times in the footpoint of the cool loops in $\beta$-$\gamma$ than $\beta$ type ARs. In conclusion, both aspects justify higher activity levels in the loop footpoints of $\beta${--}$\gamma$ ARs than $\beta$ ARs.\\


The line with high oscillator strength is affected the most due to opacity, and therefore, we do observe the reduced ratio, i.e., less than 2.0. The magnetic reconnection triggers various dynamic events within the lower solar atmosphere, like explosive events, and small-scale brightening (e.g., \citealt{Isobe2007, Peter2014, Gupta2015, Kayshap2017}). Such events can enhance the electron density along the line of sight (LOS), and as a result, a thick atmosphere will form. The reabsorption probability of Si~{\sc iv} 1393.78{\AA} photon in the thick atmosphere is twice in comparison to that of Si~{\sc iv} 1402.77~{\AA}. Hence, the intensity of Si~{\sc iv} 1393.78{\AA} reduces more and more as electron density increases in the LOS, and as a result, the ratio reduces more and more. In the present study, 23\% locations in $\beta$ ARs and 31\% locations in $\beta${--}$\gamma$ ARs have ratio values less than 1.90 (i.e., significantly less than 2), and we know that footpoints of the cool loops are much prone area to the dynamic events mentioned above (e.g., \citealt{Huang2015, Huang2019, Rao2019a, Rao2019b, Srivastava2020}). In addition, we have seen complex spectral profiles in the footpoints of the cool loop which are most probably due to the occurrence of magnetic reconnection in the vicinity of footpoints. Hence, such events at the loop footpoints have enhanced the opacity, and it is reflected as the reduced ratio of Si~{\sc iv} lines. On the other hand, 29\% locations in $\beta$ ARs and 24\% locations in $\beta${--}$\gamma$ ARs have an intensity ratio greater than 2.10, and the intensity ratio higher than 2 is due to the resonant scattering (e.g., \cite{Wood2000, Gonti2018, DT2020}). Hence, it may also be possible that the resonant scattering is happening in the vicinity of footpoints. While further studies based on the modelling of Si~{\sc iv} lines are required to know this issue in great detail.\\

The estimation of the chi-square value is necessary to know the level of complexity in the spectral profiles. The chi-square analysis establishes that $\beta$-$\gamma$ type ARs have more than 26 times more complex profiles than the complex profiles of $\beta$ type ARs. Hence, we can say that the footpoints of cool loops are much more complex in $\beta$-$\gamma$ type ARs than that in $\beta$ type ARs. In addition, we have also noticed complex profiles (i.e., non-Gaussian, broad, and multi-peak profiles) within the footpoints of cool loops of both ARs and similar observational findings are already reported (e.g., \citealt{Huang2015, Rao2019b, Srivastava2020}).\\

In conclusion, we find that the footpoints of the cool loop in $\beta$ $\gamma$ ARs are far more complex than that in $\beta$ type ARs, and the intensity $\&$ FWHM are higher in $\beta${--}$\gamma$ ARs than those in $\beta$ type ARs. On the contrary, the Doppler velocities are the same in both ARs. 

\section*{Acknowledgements}
We gratefully acknowledge the reviewer (Dr. Jaroslav Dud{\'\i}k) for his constructive comments that substantially improved the manuscript. B. Suresh Babu and P. Kayshap express gratitude and acknowledgment to Dr. Georgios Chintzoglou (LMSAL)and Dr. Bart De Pontieu (LMSAL) for the valuable inputs during a discussion at Hinode-16/IRIS-13 meeting. Based on their suggestions, we modified the analysis, particularly for the Si~{\sc iv} ratio estimation. P. Jel{\'\i}nek acknowledges support from grant 21-16508J of the Grant Agency of the Czech Republic. IRIS is a NASA small explorer mission developed and operated by LMSAL with mission operations executed at NASA Ames Research Center and major contributions to downlink communications funded by ESA and the Norwegian Space Centre. We also acknowledge the use of HMI/SDO. 
\section*{Data Availability}
The data underlying this article are available at \url{https://iris.lmsal.com/data.html} (NASA/IRISwebsite) and at \url{https://iris.lmsal.com/search/}(LMSAL search website). Note that IRIS data are publicly available with the observation ID OBS 3630088076, OBS 3620106076, OBS 3620108077, OBS 3840257196, OBS 3610108077, OBS 3860257446, OBS 3620110077, OBS 3620110077, OBS 3620108077, OBS 3610108077. 
\bibliographystyle{mnras}
\bibliography{reference.bib} 




\appendix

\section{Nature of Spectral Profiles}
\label{sect:appendix}
In this appendix, we have displayed some spectral profiles from the footpoints of cool loops of $\beta$ AR11937 (see panels (a) to (d); Figure~\ref{fig:fig9}) and  $\beta$-$\gamma$ type AR 11934 (see panels (e) to (h); Figure~\ref{fig:fig9}).  
\begin{figure*}
    \includegraphics[trim=4.0cm 2.2cm 2.5cm 1.0cm, scale=0.8]{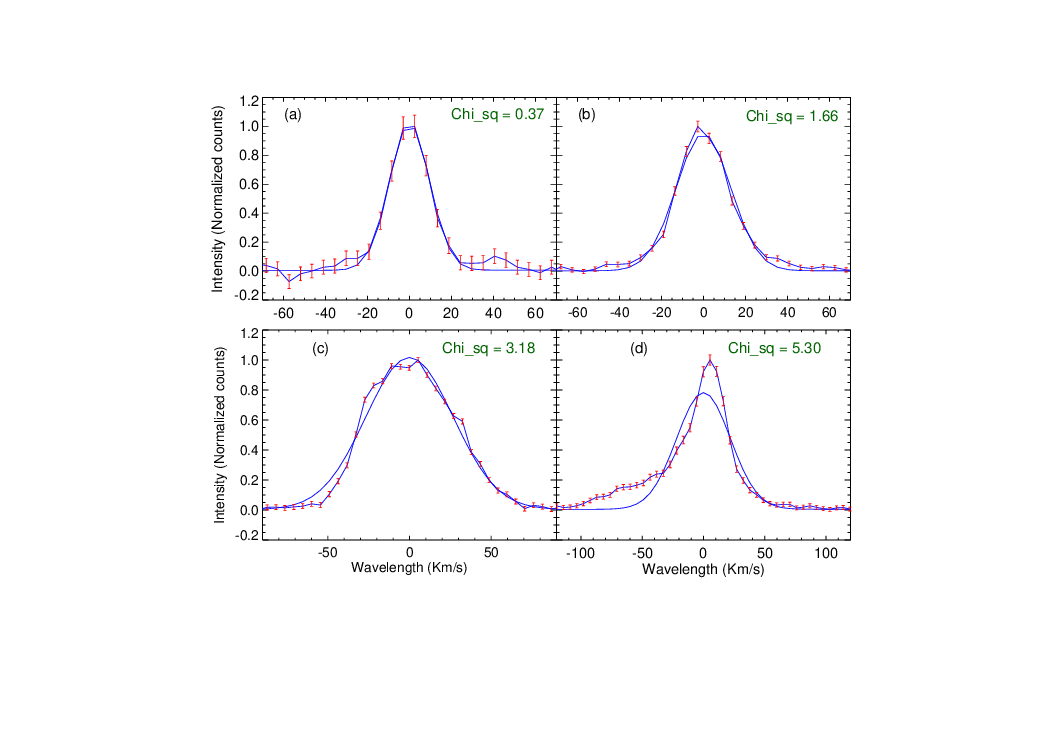}
    \includegraphics[trim=4.0cm 2.2cm 4.5cm 8.0cm, scale=0.8]{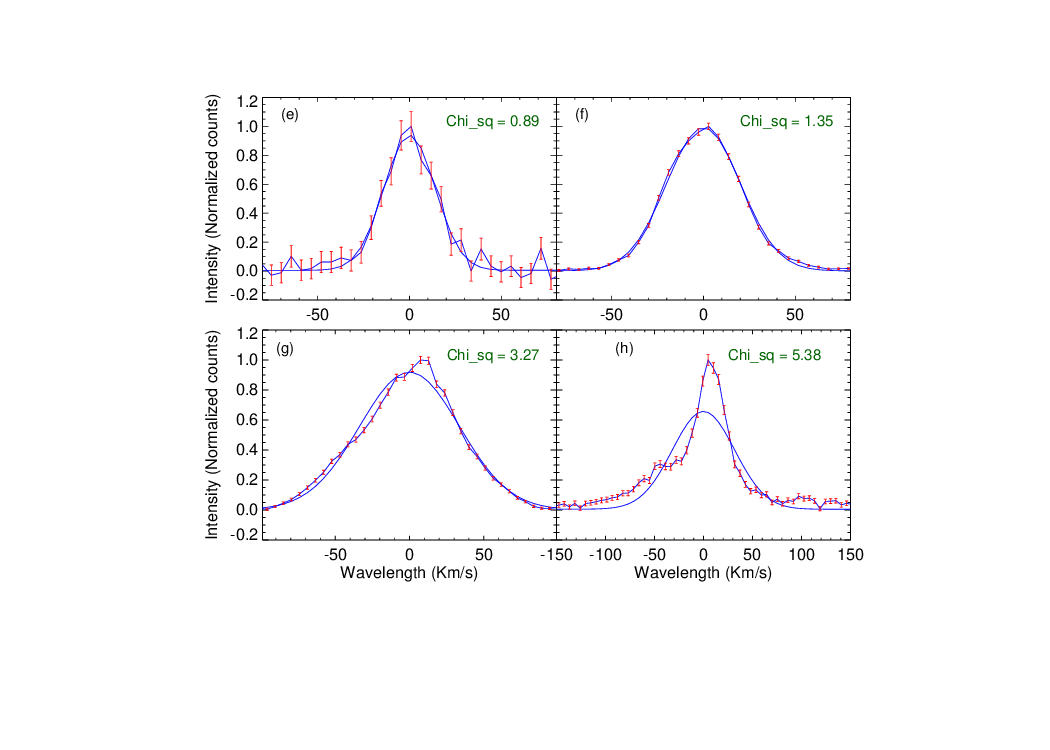}
   \caption{The panels (a) to (d) and panels (e) to (h)) show spectral profiles (black curve) along with a single Gaussian fit (blue curve) from the first $\beta$ AR and $\beta$-$\gamma$ AR, respectively. The profiles are well fitted with single Gaussian if the chi-square value is less than 3 (see panels (a), (b), (e), and (f)). The spectral profile starts to deviate from single Gaussian when the chi-square value goes beyond 3 (see panels ((c) and (g)). The profiles completely deviate from single Gaussian if the ch-square value is much higher than 3 (see panels (d) and (h).}
    \label{fig:figA1}
\end{figure*}
In each panel, the black solid line displays the observed spectra with corresponding errors in red color. The overplotted blue color line is the single Gaussian fit on the observed spectra. The reduced chi-square values are mentioned in each panel. \\
The observed spectra are well fitted by single Gaussian when the reduced chi-square values are less than 1 (see panels (a) $\&$ (b) {--} profiles from $\beta$ AR, and panels (e) $\&$ from $\beta$ $\gamma$ AR). Although, the single Gaussian fits deviate as the chi-square value exceeds 3 (see panels (c) and (g)). Lastly, we mention that if the reduced chi-square value is much higher than 3 then single Gaussian, significantly, deviates from the observed spectra. Manually, we have checked several profiles from both types of AR, and it is found that the spectral profiles have chi-square values less than 3 can be characterized by a single Gaussian. Therefore, we have considered chi-square 3 as the threshold value. 


\section{Mean Calibrated Intensities of all Boxes at loop Footpoint} \label{sect:appendix_mean_intensity}

In this appendix, we have shown the mean calibrated intensities of all selected boxes in all five $\beta$ and $\beta${--}$\gamma$ ARs, namely, table~\ref{tab:mean_intensity1} for the first $\beta$ (i.e., AR11937) and first $\beta${--}$\gamma$ AR (i.e., AR11934), table~\ref{tab:mean_intensity2} for second $\beta$ (i.e., AR12641) and second $\beta${--}$\gamma$ AR (i.e., AR12712), table~\ref{tab:mean_intensity3} for third $\beta$ (i.e., AR12832) and third $\beta${--}$\gamma$ AR (i.e., AR13226), table~\ref{tab:mean_intensity4} for fourth $\beta$ (i.e., AR12458) and fourth $\beta${--}$\gamma$ AR (i.e., AR10317), and table~\ref{tab:mean_intensity5} for fifth $\beta$ (i.e., AR12692) and fifth $\beta${--}$\gamma$ AR (i.e., AR13180). 
\begin{table}

 \caption{Mean Calibrated Intensity values of all box for first $\beta$ (left column) and first $\beta${--}$\gamma$ AR (right column). }
	\label{tab:mean_intensity1}
	\begin{tabular}{lcr} 
 
		\hline
    Box. No. &  Mean Intensity\\
		\hline
  
		b1 & 94.82\\
  \hline
		b2 & 86.48\\
  \hline
		b3 & 86.08\\
  \hline
  	    b4 & 80.39\\
  \hline
  	  b5 & 51.19\\
  \hline
            b6 & 102.74\\	  
  \hline
    	  b7 & 125.97\\
  \hline
    	  b8 & 75.17\\
  \hline
    	  b9 & 24.66\\
  \hline
    	  b10 & 40.29\\
  \hline
      	b11 & 69.11\\
  \hline
      	b12 & 70.65\\
  \hline
      	b13 & 46.00\\
  \hline
     	  b14 & 23.31\\
  \hline
     	  b15 & 36.87\\
  \hline
     	  b16 & 75.92\\
  \hline
	\end{tabular}
\begin{tabular}{lcr} 
		\hline
    Box. No. &  Mean Intensity\\
		\hline
  
		b1 & 751.34\\
  \hline
		b2 & 1650.98\\
  \hline
		b3 & 150.11\\
  \hline
  	    b4 & 246.08\\
  \hline
  	  b5 & 282.70\\
  \hline
            b6 & 177.18\\	  
  \hline
    	  b7 & 204.71\\
  \hline
    	  b8 & 552.47\\
  \hline
    	  b9 & 244.19\\
  \hline
    	  b10 & 130.35\\
  \hline
      	b11 & 92.00\\
  \hline
      	b12 & 103.23\\
  \hline
      	b13 & 58.40\\
  \hline
     	  b14 & 630.00\\
  \hline
     	  b15 & 39.63\\
  \hline
     	  b16 & 61.10\\
  \hline
            b17 & 19.60\\
  \hline
            b18 & 130.23\\
  \hline
            b19 & 47.75\\
  \hline
            b20 & 51.85\\
  \hline
	\end{tabular}
\end{table}
\medskip
\begin{table}

 \caption{Same as table~\ref{tab:mean_intensity1} but for second $\beta$ (left column) and second $\beta${--}$\gamma$ AR (right column). }   
	\label{tab:mean_intensity2}
	\begin{tabular}{lcr} 
 
		\hline
    Box. No. &  Mean Intensity\\
		\hline
  
		b1 & 90.18\\
  \hline
		b2 & 58.29\\
  \hline
		b3 & 112.49\\
  \hline
  	    b4 & 85.82\\
  \hline
  	  b5 & 93.01\\
  \hline
            b6 & 90.64\\	  
  \hline
    	  b7 & 139.50\\
  \hline
    	  b8 & 100.29\\
  \hline
    	  b9 & 53.81\\
  \hline
    	  b10 & 51.68\\
  \hline
      	b11 & 49.48\\
  \hline
      	b12 & 65.52\\
  \hline
      	b13 & 68.20\\
  \hline
     	  b14 & 32.77\\
  \hline
     	  b15 & 31.19\\
  \hline
     	  b16 & 35.89\\
  \hline
	\end{tabular}
\begin{tabular}{lcr} 
		\hline
    Box. No. &  Mean Intensity\\
		\hline
  
		b1 & 144.37\\
  \hline
		b2 & 97.69\\
  \hline
		b3 & 54.65\\
  \hline
  	    b4 & 53.03\\
  \hline
  	  b5 & 190.97\\
  \hline
            b6 & 128.44\\	  
  \hline
    	  b7 & 98.93\\
  \hline
    	  b8 & 81.64\\
  \hline
    	  b9 & 46.18\\
  \hline
    	  b10 & 37.66\\
  \hline
      	b11 & 86.30\\
  \hline
      	b12 & 65.27\\
  \hline

	\end{tabular}
\end{table}
\begin{table}

 \caption{Same as table~\ref{tab:mean_intensity1} but for third $\beta$ (left column) and third $\beta${--}$\gamma$ AR (right column).)}
	\label{tab:mean_intensity3}
	\begin{tabular}{lcr} 
 
		\hline
    Box. No. &  Mean Intensity\\
		\hline
  
		b1 & 43.93\\
  \hline
		b2 & 57.44\\
  \hline
		b3 & 31.64\\
  \hline
  	    b4 & 76.18\\
  \hline
  	  b5 & 72.22\\
  \hline
            b6 & 33.27\\	  
  \hline
    	  b7 & 51.25\\
  \hline
    	  b8 & 36.83\\
  \hline
    	  b9 & 37.30\\
  \hline
    	  b10 & 40.41\\
  \hline
      	b11 & 24.20\\
  \hline
      	b12 & 22.67\\
  \hline
      	b13 & 66.67\\
  \hline
	\end{tabular}
\begin{tabular}{lcr} 
		\hline
    Box. No. &  Mean Intensity\\
		\hline
  
		b1 & 302.71\\
  \hline
		b2 & 154.46\\
  \hline
		b3 & 771.70\\
  \hline
  	    b4 & 227.58\\
  \hline
  	  b5 & 131.70\\
  \hline
            b6 & 417.15\\	  
  \hline
    	  b7 & 106.17\\
  \hline
    	  b8 & 151.48\\
  \hline
    	  b9 & 968.28\\
  \hline
    	  b10 & 148.63\\
  \hline
      	b11 & 128.41\\
  \hline
      	b12 & 59.28\\
  \hline
      	b13 & 139.02\\
  \hline
     	  b14 & 156.00\\
  \hline
     	  b15 & 94.91\\
  \hline
     	  b16 & 66.96\\
  \hline
            b17 & 80.42\\
  \hline
            b18 & 73.92\\
  \hline
            b19 & 127.94\\
  \hline
            b20 & 64.22\\
  \hline
            b21 & 141.44\\
  \hline
            b22 & 160.10\\
  \hline
            b23 & 114.51\\
  \hline
	\end{tabular}
\end{table}
\medskip
\begin{table}

 \caption{Same as table~\ref{tab:mean_intensity1} but for fourth $\beta$ (left column) and fourth $\beta${--}$\gamma$ AR (right column).}   
	\label{tab:mean_intensity4}
	\begin{tabular}{lcr} 
 
		\hline
    Box. No. &  Mean Intensity\\
		\hline
  
		b1 & 25.11\\
  \hline
		b2 & 23.81\\
  \hline
		b3 & 37.84\\
  \hline
  	    b4 & 53.62\\
  \hline
  	  b5 & 65.02\\
  \hline
            b6 & 54.71\\	  
  \hline
    	  b7 & 19.59\\
  \hline
    	  b8 & 39.95\\
  \hline
    	  b9 & 33.21\\
  \hline
    	  b10 & 39.52\\
  \hline
      	b11 & 35.04\\
  \hline
      	b12 & 47.43\\
  \hline
      	b13 & 14.07\\
  \hline
            b14 & 21.38\\
  \hline
		b15 & 24.10\\
  \hline
		b16 & 46.52\\
  \hline
  	    b17 & 41.84\\
  \hline
  	  b18 & 29.54\\
  \hline
            b19 & 20.36\\	  
  \hline
    	  b20 & 19.99\\
  \hline
    	  b21 & 65.40\\
  \hline
    	  b22 & 27.38\\
  \hline
    	  b23 & 20.42\\
  \hline
      	b24 & 22.97\\
  \hline
      	b25 & 12.30\\
  \hline
	\end{tabular}
\begin{tabular}{lcr} 
		\hline
    Box. No. &  Mean Intensity\\
		\hline
  
		b1 & 152.28\\
  \hline
		b2 & 271.07\\
  \hline
		b3 & 93.75\\
  \hline
  	    b4 & 168.74\\
  \hline
  	  b5 & 282.74\\
  \hline
            b6 & 153.67\\	  
  \hline
    	  b7 & 197.77\\
  \hline
    	  b8 & 100.07\\
  \hline
    	  b9 & 140.35\\
  \hline
    	  b10 & 149.20\\
  \hline
      	b11 & 110.04\\
  \hline
      	b12 & 142.64\\
  \hline

	\end{tabular}
\end{table}
\begin{table}

 \caption{Same as table~\ref{tab:mean_intensity1} but for fifth $\beta$ (left $\&$ middle columns) and fifth $\beta${--}$\gamma$ AR (right column).}   
	\label{tab:mean_intensity5}
	\begin{tabular}{lcr} 
 
		\hline
    Box. No. &  Mean Intensity\\
		\hline
  
		b1 & 87.12\\
  \hline
		b2 & 66.12\\
  \hline
		b3 & 123.33\\
  \hline
  	    b4 & 89.60\\
  \hline
  	  b5 & 135.68\\
  \hline
            b6 & 76.61\\	  
  \hline
    	  b7 & 118.24\\
  \hline
    	  b8 & 67.19\\
  \hline
    	  b9 & 90.48\\
  \hline
    	  b10 & 57.24\\
  \hline
      	b11 & 72.43\\
  \hline
      	b12 & 52.39\\
  \hline
      	b13 & 124.57\\
  \hline
            b14 & 87.52\\
  \hline
		b15 & 74.63\\
  \hline
		b16 & 43.38\\
  \hline
  	    b17 & 58.50\\
  \hline
  	  b18 & 53.38\\
  \hline
            b19 & 70.0\\	  
  \hline
    	  b20 & 52.14\\
  \hline
    	  b21 & 109.29\\
  \hline
    	  b22 & 34.76\\
  \hline
    	  b23 & 91.69\\
  \hline
      	b24 & 42.82\\
  \hline
      	b25 & 48.14\\
  \hline
       \end{tabular}
\begin{tabular}{lcr} 
 
		\hline
    Box. No. &  Mean Intensity\\
  \hline
            b26 & 47.87\\
  \hline
		b27 & 83.46\\
  \hline
		b28 & 47.41\\
  \hline
  	    b29 & 95.81\\
  \hline
  	  b30 & 126.27\\
  \hline
            b31 & 67.35\\	  
  \hline
    	  b32 & 123.55\\
  \hline
    	  b33 & 128.52\\
  \hline
    	  b34 & 194.62\\
  \hline
    	  b35 & 46.15\\
  \hline
      	b36 & 80.54\\
  \hline
      	b37 & 29.00\\
  \hline
      	b38 & 54.17\\
  \hline
            b39 & 60.03\\
  \hline
		b40 & 77.69\\
  \hline
		b41 & 79.53\\
  \hline
  	    b42 & 96.69\\
  \hline
  	  b43 & 92.05\\
  \hline
            b44 & 117.48\\	  
  \hline
    	  b45 & 104.87\\
  \hline
    	  b46 & 70.97\\
  \hline
    	  b47 & 66.22\\
  \hline
    	  b48 & 78.14\\
  \hline
      	b49 & 65.55\\
  \hline
      	b50 & 59.51\\
  \hline
	\end{tabular}
 \end{table}
\begin{table}
\begin{tabular}{lcr} 
         &  \\
         & \\
         & \\

		\hline
    Box. No. &  Mean Intensity\\
		\hline
  
		b1 & 125.44\\
  \hline
		b2 & 187.04\\
  \hline
		b3 & 124.65\\
  \hline
  	    b4 & 170.00\\
  \hline
  	  b5 & 122.18\\
  \hline
            b6 & 98.70\\	  
  \hline
    	  b7 & 81.47\\
  \hline
    	  b8 & 51.07\\
  \hline
    	  b9 & 125.14\\
  \hline
    	  b10 & 58.44\\
  \hline
      	b11 & 71.00\\
  \hline
      	b12 & 33.51\\
  \hline
      	b13 & 29.88\\
  \hline

	\end{tabular}
\end{table}

\bsp	
\label{lastpage}

\end{document}